\documentstyle[12pt,floats,aps,bezier,twoside,epsfig,epsf]{revtex}
\newcommand {\beq}{\begin{eqnarray}}
\newcommand {\eeq}{\end{eqnarray}}
\newcommand {\non}{\nonumber\\}

\newcommand {\1}[1]{\frac{1}{#1}}

\newcommand {\thb}{\bar{\theta}}

\newcommand {\ph}{\varphi}

\newcommand {\sig}{\sigma}

\newcommand {\del}{\partial}
\newcommand {\dagg}{^{\dagger}}

\newcommand {\lam}{\lambda}

\newcommand {\tr}{{\rm tr}\,}

\newcommand{\hs}[1]{\hspace{#1 mm}}

\newcommand{\Lam}{\Lambda}

\newcommand{\kahler}{K\"ahler }

\begin{document}

\thispagestyle{empty}

\setcounter{page}{1}

\title {Massive Hyper-K{\"a}hler Sigma Models and
 BPS Domain Walls}

\author{
\normalsize
  {\large \bf 
  Masato~Arai~$^{a}$}
\footnote{\it e-mail address: 
arai@fzu.cz},  
 {\large \bf 
Muneto~Nitta~$^{b}$}
\footnote{\it e-mail address: 
nitta@physics.purdue.edu}
\footnote{Address after December 11th, Tokyo Institute of Technology.
}
~and~~  {\large \bf 
Norisuke~Sakai~$^{c}$}
\footnote{\it e-mail address: 
nsakai@th.phys.titech.ac.jp}}
\maketitle
\vskip 1.5em
\begin{center}
{\it $^{a}$Institute of Physics, AS CR, 
  182 21, Praha 8, Czech Republic \\
$^{b}$Department of Physics, Purdue University, 
West Lafayette, IN 47907-1396, USA\\
and \\
  $^{c}$
Department of Physics, Tokyo Institute of 
 Technology \\
Tokyo 152-8551, JAPAN 
}
\end{center}

With the non-Abelian Hyper-K{\"a}hler quotient by $U(M)$ and $SU(M)$ 
 gauge groups, we give 
 the massive Hyper-K{\"a}hler sigma models that are not toric in the 
 ${\cal N}=1$ superfield formalism.
The $U(M)$ quotient gives $N!/[M! (N-M)!]$ 
 ($N$ is a number of flavors) discrete vacua 
 that may allow various types of domain walls, 
 whereas the $SU(M)$ quotient gives no discrete vacua. 
We derive BPS domain wall solution in the case of 
 $N=2$ and $M=1$ in the $U(M)$ quotient model.

\bigskip

\section{Introduction}
It is well known that topological solutions are of importance
 in various areas of particle physics. 
Recently, there was renewed interest in such solutions because of their
 crucial role in the brane world scenario \cite{LED,RS}. 
In this brane-world scenario, our four-dimensional world 
 is to be realized 
 on topological objects like domain walls or brane-junctions. 
Supersymmetry (SUSY) can be implemented in these models, and
 it is actually a powerful device for constructing their 
 topological solutions.
Viewing the four-dimensional world as a domain wall, we are led to deal
 with SUSY theories with eight supercharges in five dimensions.

SUSY with eight supercharges is very restrictive. 
In theories involving only massless scalar multiplets (hypermultiplets),
 non-trivial interactions can only arise from nonlinearities in kinetic
 term, say nonlinear sigma models (NLSMs).
Prior to studying the genuine five-dimensional theories with hypermultiplets,
 it is instructive to start with similar SUSY theories in four
 dimensions, i.e., ${\cal N}=2,~d=4$ theories.
The analysis of the four-dimensional theory could then be of help in
 studying the brane world scenario based on SUSY theories in higher 
 dimensions\cite{AFNS}.

With regard to rigid ${\cal N}=2$ SUSY the target manifold of
 the hypermultiplet $d=4$ NLSMs
 must be Hyper-K{\"a}hler (HK) \cite{AF1}.
In these theories, the scalar potential can be obtained only if
 the hypermultiplets acquire masses by the Scherk-Schwarz mechanism
 \cite{SS} because of the appearance of central
 charges in the ${\cal N}=2$ Poincar\'e superalgebra \cite{AF2}.
The NLSMs with the scalar potential in ${\cal N}=2$
 theories are called the massive HK NLSMs. 

A large class of HK manifold is given by toric HK manifolds that are
 defined as HK manifolds of real dimension $4n$ admitting mutually
 commuting $n$ Abelian tri-holomorphic isometries.
In the massive HK NLSMs on toric HK manifolds,
 many interesting BPS solitons were constructed 
 in the component formalism \cite{AT,GTT1,GTT2,others}
 as well as off-shell formulation \cite{To,ANNS,AIN}. 
The potential term of 
 the massive $T^*{\bf C}P^{N-1}$ model which is toric
 comes from the mass terms of the 
 hypermultiplets when the NLSM is 
 constructed as the quotient by the 
 $U(1)$ gauge group~\cite{To,ANNS}. 
We call this formulation of massive HK NLSMs 
 as ``the massive HK quotient method", 
 since massless case is just a HK quotient found 
 in Refs.~\cite{LR,HKLR1}. 
One of the advantages of our massive HK quotient 
 is that the off-shell formulation of the SUSY 
 NLSMs is possible~\cite{ANNS}. 
Off-shell formulation is powerful to extend the models
to those with other isometries, and/or gauge symmetries 
and to those coupled with gravity, since (part of) SUSY is manifest.
Any {\sl toric} HK manifolds can be constructed 
 using an {\sl Abelian} HK quotient~\cite{GGR,toric}.
Therefore an off-shell formulation of general 
 massive toric HK NLSMs~\cite{GTT1} 
 can be obtained 
 using the massive HK quotient with the 
Abelian gauge theories. 
On the other hand, 
 a massless HK NLSM other than 
 the toric HK target manifolds has been 
 obtained as a quotient using non-Abelian gauge group 
 by Lindstr\"om and Ro\v{c}ek~\cite{LR} for {\sl massless} 
 case only (without potential terms). 

In this talk, we discuss massive NLSMs in ${\cal N}=2,~d=4$ theories and
 their BPS domain wall solutions.
With HK quotient method, massive NLSMs on 
 cotangent bundles over the Grassmann manifolds, 
 $T^*G_{N,M}$, which are not toric, are 
 obtained along with their generalization. 
These models are constructed in ${\cal N}=1$ superfield formalism.
BPS domain wall solution is given in the simplest case,  
 the Eguchi-Hanson target manifold \cite{EH} ($N=2$ and $M=1$).
This talk is based on our papers \cite{ANNS,ANS} in which
 analysis by a fully off-shell ${\cal N}=2$ superspace 
 (the harmonic superspace \cite{HSS})
 formalism is also discussed in detail.

\section{Massive HK quotient by $U(M)$ gauge group}
We consider ${\cal N}=2$ SUSY QCD with 
 $N$-flavors and a $U(M)$ gauge group.
In terms of ${\cal N}=1$ superfields, 
 ${\cal N} =2$, $NM$ hypermultiplets
 can be decomposed 
into $(N \times M)$- and $(M \times N)$-matrix 
 chiral superfields $\Phi(x,\theta,\thb)$ 
 and $\Psi(x,\theta,\thb)$, and 
 ${\cal N}=2$ vector multiplets for the 
 $U(M)$ gauge symmetry 
 can be decomposed into $M\times M$ matrices of 
${\cal N}=1$ vector superfields $V=V^A(x,\theta,\thb) T_A$ 
 and chiral superfields 
 $\Sigma = \Sigma^A (x,\theta,\thb) T_A$,  
 with $M\times M$ matrices $T_A~(A=1,\dots,M)$ of the fundamental 
 representation of the generators of the $U(M)$ gauge group.
In order that the vector multiplets are treated as Lagrange multipliers,
 we take strong coupling limit $g\rightarrow \infty$, and
 drop the kinetic term.
The gauge invariant Lagrangian is given by
\beq
&& {\cal L} = \int d^4 \theta
 \left[ \tr (\Phi\dagg\Phi e^V )  
 + \tr (\Psi\Psi\dagg e^{-V}) - c\, \tr V \right]  \non
&&\hs{5} + \left[ \int d^2\theta \,
       \left(\tr \left\{ \Sigma (\Psi \Phi - b {\bf 1}_M) \right\} 
      + \sum_{a=1}^{N-1} m_a \tr (\Psi H_a \Phi)\right) 
         + {\rm c.c.}\right] ,
\label{linear}
\eeq
where we have absorbed a common mass 
 of hypermultiplets into the field $\Sigma$, 
 denoted $m_a$ ($a=1,\cdots,N-1$) 
 as complex mass parameters 
 and $H_a$ are diagonal traceless matrices,  
 interpreted as the Cartan generators of $SU(N)$ below. 
The electric and magnetic Fayet-Iliopoulos (FI) parameters 
 are denoted as $c\in {\bf R}$ and $b \in {\bf C}$, respectively.  
Note that $U(M)$ gauge symmetry is complexified.

Next we eliminate the auxiliary superfields 
 $V$ and $\Sigma$ in the superfield formalism. 
Their equations of motion read from 
 Eq.(\ref{linear}):
\begin{eqnarray}
 && {\partial {\cal L} \over \partial V} 
 = \Phi\dagg\Phi e^V - e^{-V} \Psi\Psi\dagg - c {\bf 1}_M = 0 \; , 
   \label{EOM-V}\\
 && {\partial {\cal L} \over \partial \Sigma} 
 = \Psi \Phi - b {\bf 1}_M = 0\;.
   \label{EOM-sig}
\end{eqnarray}
From the first equation, $V$ can be solved 
\beq
 e^V = {c\over 2} (\Phi\dagg\Phi)^{-1} 
  \left({\bf 1}_M  \pm \sqrt {{\bf 1}_M 
     + {4\over c^2} \Phi\dagg\Phi \Psi\Psi\dagg}\right) \,. 
  \label{sol-V}
\eeq 
Substituting this back into (\ref{linear}), 
 we obtain the K\"ahler potential for 
 the Lindstr\"{o}m-Ro\v{c}ek metric~\cite{LR} 
\begin{equation}
 K = c\, \tr \sqrt{{\bf 1}_M + {4\over c^2} \Phi\dagg\Phi \Psi\Psi\dagg} 
   - c\, \tr \log \left( {\bf 1}_M 
    + \sqrt{{\bf 1}_M + {4\over c^2} \Phi\dagg\Phi \Psi\Psi\dagg}\right)
   + c\, \tr \log \Phi\dagg\Phi  \;. \label{kahler}
\end{equation}

Fixing the complexified $U(M)$ gauge symmetry 
 and solving constraint (\ref{EOM-sig}), 
we obtain the Lagrangian of the NLSM
 in terms of independent superfields. 
To this end we should consider two cases 
 i) $b = 0$ and ii) $b \neq 0$ separately.\\ 
 i) $b = 0$. In this case, a gauge can be fixed as
\beq
 \Phi = \left(\begin{array}{c}
         {\bf 1}_M \cr
         \varphi
	\end{array}
        \right),\hs{5}
        \Psi = (- \psi\ph, \psi) \;, \label{fixing1}          
\eeq
 with $\ph$ and $\psi$ being 
 $[(N-M)\times M]$- and $[M\times (N-M)]$-matrix chiral 
 superfields, respectively. 
The superpotential becomes
\beq
 W = \sum_a m_a \tr \left[
   (- \psi\ph, \psi) H_a \left(
                         \begin{array}{c}
                           {\bf 1}_M \cr \ph
                         \end{array}\right) 
  \right]  
  = \sum_a m_a \tr \left[ H_a
    \left(
    \begin{array}{cc}
          -\psi\ph & \psi \cr 
          - \ph\psi\ph & \ph\psi
    \end{array}
    \right)
      \right]\,. 
          \label{superpot1}
\eeq
\\
ii) $b\neq 0$. 
In this case, we can take a gauge as~\cite{LR}
\beq
 \Phi = \left(
        \begin{array}{c}
          {\bf 1}_M \cr \ph
        \end{array}
        \right)  
        Q \;, \hs{5} 
 \Psi = Q ({\bf 1}_M, \psi) \;, \hs{5}
 Q = \sqrt b ({\bf 1}_M + \psi\ph)^{-\1{2}} \;, 
 \label{fixing2}
\eeq
with $\ph$ and $\psi$ being again
 $[(N-M)\times M]$- and $[M\times (N-M)]$-matrix chiral 
 superfields, respectively. 
In this case, the superpotential is given by 
\beq
 W = b \sum_a m_a \tr \left[
    H_a \left(\begin{array}{c}
          {\bf 1}_M \cr \ph
        \end{array}\right) 
    ({\bf 1}_M + \psi\ph)^{-1}
    ({\bf 1}_M, \psi)  
   \right] \;. \label{superpot2}
\eeq
These two cases are not holomorphically transformed to 
 each other, because they make different complex structures 
 manifest.

\medskip
We can find the bundle structure of the manifold as follows: 
i) $b=0$. 
Putting $\psi=0$, the \kahler potential becomes 
\beq
 K|_{\psi=0} = c\, \tr \log (1+\ph\dagg\ph) \;, 
\eeq
which is the one of the Grassmann manifold. 
Therefore $\ph$ parameterize the base Grassmann manifold, 
 whereas $\psi$ the cotangent space as the fiber, 
 with the total space being the cotangent bundle 
 over the Grassmann manifold $T^* G_{N,M}$. 
ii) $b\neq 0$. 
In the case of $T^*{\bf C}P^{N-1}$ of $M=1$, 
 the base manifold is embedded by 
 $\ph = \psi\dagg$~\cite{AF3}.\footnote{
 This embedding $\ph = \psi\dagg$ should hold for 
 a matrix of general $M$, 
 although we have not proved it yet.
}

There exists the manifest duality between 
two theories with 
$U(M)$ gauge and $U(N-M)$ gauge symmetries 
and the same flavor $SU(N)$ symmetry. 
This comes directly from the duality in 
the base Grassmann manifold $G_{N,M} \simeq G_{N,N-M}$. 

For $M=1$ ($M=N-1$) namely  
 for the $U(1)$ [$U(N-1)$] gauge symmetry,
 this model reduces to 
 $T^* {\bf C}P^{N-1} \simeq T^* G_{N,1} 
 (\simeq T^* G_{N,N-1})$ \cite{Ca} 
 which we discussed in detail in \cite{ANNS}. 
Moreover if $N=2$ the manifold $T^*{\bf C}P^1$ 
 is the Eguchi-Hanson space.
A nontrivial model in the lowest dimensions 
 other than $T^*{\bf C}P^{N-1}$
 is the case of $N=4, M=2$. 
The manifold is 
 $T^* G_{4,2} = T^* [SU(4)/ SU(2) \times SU(2) \times U(1)] 
 = T^* [SO(6)/SO(4) \times U(1)] \equiv T^* Q^4$
 in which the base manifold $Q^4$ 
 is called the Klein quadric space.

\section{Vacuum structure}
\subsection{Vacua in the massive $T^* {\bf C}P^{N-1}$ model}
In this subsection we discuss 
 $T^*{\bf C}P^{N-1} = T^* G_{N,1}$ of $M=1$. 
Without loss of generality we consider the case of 
 $b=0$ and $c\neq 0$. 
The dynamical matrix fields are 
 column and row vectors like 
 $\ph^T = (\ph^1,\cdots,\ph^{N-1})$ and 
 $\psi = (\psi^1,\cdots,\psi^{N-1})$.

The superpotential given in (\ref{superpot1}) becomes
\beq
 W = \sum_a m_a \tr \left[H_a
       \left(
       \begin{array}{cc}
           - \psi\cdot \ph & \psi \cr
          - \ph (\psi \cdot \ph) & \ph \otimes \psi
       \end{array}
       \right)  
      \right] \;.
\eeq
We take $H_a$ $(a=1,\cdots ,N-1)$ as 
\beq
 H_a =  \1{\sqrt{a(a+1)}} 
        \;{\rm diag.}\; (1,\cdots,1, -a, 0,\cdots,0)\;,
\eeq
where $-a$ is the ($a+1$)-th component, 
 with a normalization given by 
 the trace $\tr (H_a H_b) = \delta_{ab}$.
Then the superpotential can be calculated as
\beq
 W = - \sum_a M_a \psi^a \ph^a \;, \hs{5}
 M_a \equiv \sqrt{a \over a+1} m_a 
           + \sum_{b=1}^a {m_b \over \sqrt{b(b+1)}} \;. 
 \label{superpotCP}
\eeq
Therefore the derivatives of $W$ with respect to fields are
\beq
  \del_{\ph^a} W = - M_a \psi^a \;, \hs{5}
 \del_{\psi^a} W = - M_a \ph^a \; \hs{5} (\mbox{no sum}) \;.
\eeq
These vanish only at the origin $\ph= \psi^T =0$,
 which is the only one vacuum in the regular region of 
 these coordinates because the metric is regular there.

This model, however, contains more vacua, 
 because the whole manifold is covered by the several 
 coordinate patches and the vacuum exists at the origin of each
 coordinate patch.
To see this we concentrate on the base ${\bf C}P^{N-1}$ for a while.
We consider the fields before the gauge fixing, 
 $\Phi \equiv \phi^A = (\phi^1, \cdots, \phi^N)^T$ 
 ($A= 1,\cdots,N$) 
 called the homogeneous coordinates,  
 in which we need an identification 
 by the gauge transformation 
 $\phi^A \sim e^{i\Lam} \phi^A$. 
In the region $\phi^1 \neq 0$ we can take 
 a patch $\ph^i = \phi^{i+1}/\phi^1$ $(i= 1,\cdots,N-1)$,
 which was used in Eq.~(\ref{fixing1}). 
Here let us write these coordinates as 
 $\ph_{(1)}^i = \phi^{i+1}/\phi^1$. 
In the same way, in the region of 
 $\phi^A \neq 0$, 
 we can take the $A$-th patch defined by
\beq  
 \ph_{(A)}^i 
  = \left\{\begin{array}{c}
           \phi^i/\phi^A    \hs{5} (1 \leq i \leq A-1)  \cr
           \phi^{i+1}/\phi^A \hs{5} (A \leq i \leq N-1)
    \end{array}\right.
  \;.
\eeq
We thus have $N$ sets of patches $\{\ph_{(A)}^i \}$ 
 enough to cover the whole base manifold. 
Corresponding to each patch for the base space, 
 we manifestly have an associated patch for 
 the fiber tangent space $\{\psi_{(A)}^i \}$ from 
 Eq.~(\ref{fixing1}).  
These sets of coordinates 
 $\{\ph_{(A)}^i, \psi_{(A)}^i\}$ 
 are enough to cover the whole $T^*{\bf C}P^{N-1}$.
For each patch, 
 the origin $\ph_{(A)}^i = \psi_{(A)}^i = 0$ 
 is a vacuum. 
 Therefore the number of discrete vacua for 
 the massive $T^*{\bf C}P^{N-1}$ model is $N$, 
 which was firstly found in \cite{GTT2}.

To discuss solitons like BPS walls, their junction and lumps, 
 it may be better to consider the problem 
 in one coordinate patch. 
The other vacua appear in one patch 
 as the coordinate singularities of 
 the metric in infinities of the coordinates 
 rather than the stationary points of the 
 superpotential~\cite{NNS}. 
To see this, we consider only the base ${\bf C}P^{N-1}$ once again.
We discuss how the $A$-th vacuum ($A\neq 1$) 
 in the origin of the $A$-th coordinate patch is mapped in 
 the first patch.
The $A$-th vacuum is represented by 
 $\ph_{(A)}^i = 0$ 
 or $\phi^B/\phi^A = 0$ for all $B ( \neq A)$.
In the first coordinate patch, this point is mapped 
 into an infinite point represented by 
\beq
 \ph_{(1)}^{A-1} \to \infty \;,\hs{5}  
 \ph_{(1)}^i / \ph_{(1)}^{A-1} \to 0 
   \;\; (i \neq A-1), 
\eeq
which looks like a runaway vacuum in this patch. 
Hence, the origin and $N-1$ infinities are vacua
 in each coordinate patch~\cite{NNS}. 
As a summary, 
 if we include runaway vacua, 
 one patch is enough to describe 
 soliton solutions.
However note that the terminology ``runaway" is 
 just a coordinate-dependent concept, 
 because a runaway vacuum in one coordinate patch
 is a true vacuum in the other coordinate patch.

\begin{figure}
\begin{center}
\leavevmode
  \epsfysize=7.0cm
  \epsfbox{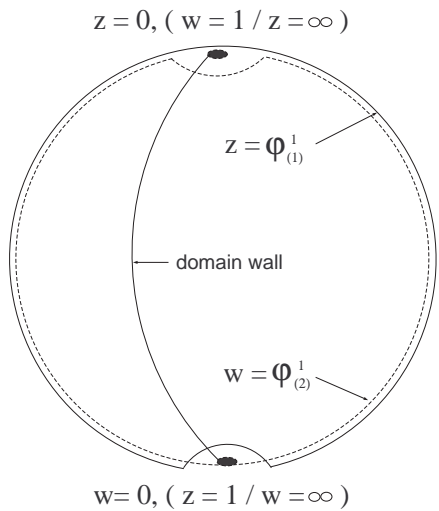} \\ 
\caption{
The base manifold of $T^*{\bf C}P^1$ and vacua.}
\label{fig1}
\end{center}
\begin{small}
Corresponding to two gauge fixing conditions, 
we have two coordinates $z$ and $w$, 
covering $S^2$ except for South (S) and North (N) Poles, 
respectively.
The origins of $z$ and $w$ ( N and S, respectively) 
are both vacua. 
The domain wall solution, 
approaching to these two vacua in spatial infinities,
is mapped to a trajectory connecting N and S in $S^2$. 
\end{small}
\end{figure}\label{CP1}

\medskip
We can also discuss the vacua without referring to 
 the local coordinate patches. 
We concentrate on the base ${\bf C}P^{N-1}$ once again. 
A point in the ${\bf C}P^{N-1}$ 
 corresponds to a complex line through the origin 
 in ${\bf C}^N$ with homogeneous coordinates $\phi^A$, 
 because of the gauge transformation 
 $\phi^A\sim e^{i\Lam}\phi^A$ as an equivalence relation.  
The first vacuum is expressed in region $\phi^1 \neq 0$ 
 by $\ph_{(1)}^i = \phi^{i+1}/\phi^1 =0$ $(i=1,\cdots,N-1)$, 
 namely $\phi^{i+1} = 0$. 
Therefore the first vacuum corresponds to 
 the $\phi^1$-axis. 
In the same way, the $A$-th vacuum corresponds to 
 the $\phi^A$-axis. 
Each vacuum is simply expressed by  
 each orthogonal axis in ${\bf C}^N$. 
Note that each axis is invariant under 
 $U(1)^{N-1}$ transformation of $H_a$ so that 
 it is a fixed point of this transformation. 

If we take $N$ orthogonal normalized basis $e_A$ 
 [with $(e_A)^*\cdot e_B = \delta_{AB}$] 
 whose components are given by 
\beq
 (e_A)^B =\delta_A^B, \label{basis}
\eeq
 a complex line in ${\bf C}^N$ can be spanned by 
 an unit vector $e' = \sum_{A=1}^N a^A e_A = U e_1$ 
 where $a^A$ is a complex number with 
 $\sum_A |a^A|^2 = 1$ and 
 $U$ is an unitary matrix $U \in U(N)$. 
Each of the $N$-vacua found above corresponds to 
 each $e_A$ ($A=1,\cdots,N$) 
 (with zero value of the cotangent space $\psi=0$).

\medskip
\underline{Example:  
 the Eguchi-Hanson space}~\cite{EH}.
The simplest model is the Eguchi-Hanson space, 
 $T^*{\bf C}P^1$ ($N=2$ and $M=1$). 
This model has two discrete vacua and admits 
 a typical domain wall solution~\cite{AT,ANNS}.
The vacua are located on the North and South Poles 
 of the base ${\bf C}P^1 \simeq S^2$ (see Fig.~\ref{fig1}).
Corresponding to two gauge fixing conditions 
 $\Phi = \left(\begin{array}{c} 1 \cr z \end{array}\right)$ 
 and $\Phi = \left(\begin{array}{c} w \cr 1 \end{array}\right)$, 
 we have two coordinate patches 
 $z \equiv \ph^1_{(1)} = \phi^2/\phi^1$ and
 $w \equiv \ph^1_{(2)} = \phi^1/\phi^2$, 
 which are related by $z = 1/w$. 
Two vacua are given by $z = 0$ and $w=0$. 
The second (first) vacuum $w=0$ ($z=0$) is mapped to 
 $z= \infty$ ($w=\infty$) in the first (second) patch, 
 which looks like a runaway vacuum.
In homogeneous coordinates, 
 these correspond to 
$\left< \Phi \right> = 
 \left(\begin{array}{c} 1 \cr 0 \end{array}\right) 
 \equiv e_1$ 
and 
$\left< \Phi \right> = 
 \left(\begin{array}{c} 0 \cr 1 \end{array} \right)
 \equiv e_2$, 
 respectively, with $\left< \Psi \right> = (0,0)$.
Also, in the coordinate independent way, 
 these two vacua correspond to the 
 $\phi^1$ and $\phi^2$ axes spanned by 
 $e_1$ and $e_2$, respectively.

Before closing this subsection, 
 we discuss the case of $b \neq 0$. 
The superpotential (\ref{superpot2}) 
 can be calculated, to give
\beq
&& W = {b \over 1+\psi\cdot\ph} 
    \left( L + \sum_{a=1}^{N-1} N_a \psi^a \ph^a \right) \,, \non
&& L \equiv \sum_{a=1}^{N-1} {m_a \over \sqrt{a(a+1)}} \;,\hs{5}
   N_a \equiv - \sqrt{a \over a+1} m_a 
  + \sum_{b=a+1}^{N-1} {m_b \over \sqrt{b(b+1)}} = L- M_a\;,
\eeq
with $M_a$ defined in (\ref{superpotCP}).
The derivatives of $W$ are
\beq
 \del_{\ph^a} W 
  &=& - {b \psi^a \over (1+ \psi\cdot\ph)^2} 
  \left[M_a - 
  \sum_{b=1}^{N-1} (M_b - M_a) \psi^b \ph^b \right]
    \; , \non
 \del_{\psi^a} W 
   &=& (\psi^a \leftrightarrow \ph^a) \;,
\eeq
 where an arrow in the second equation represents 
 the exchange of quantities in the first equation.
The origin $\ph^a=\psi^a =0$ in each patch is a vacuum. 
There is no other vacuum than these $N$ vacua;
The number of vacua should coincide with 
 the case of $b=0$ and $c\neq 0$, 
 because they are connected by the R-symmetry and 
 the physics does not depend on the difference.

\subsection{Vacua in the massive $T^* G_{N,M}$ model}
To look for vacua of the $T^* G_{N,M}$ model,
 we consider the case $b=0$ and $c\ \neq 0$ again 
 without loss of generality. 
We label the indices for the matrices as
 $\ph = (\ph_{i \alpha})$ and 
 $\psi = (\psi_{\alpha i})$ 
 in which $i=1,\cdots, N-M$ and $\alpha = 1,\cdots,M$. 
The superpotential given in Eq.~(\ref{superpot1}) 
 can be calculated as
\beq
 && W = - \sum_{\alpha =1}^M \sum_{i=1}^{N-M} 
     M_{\alpha i} \ph_{i \alpha} \psi_{\alpha i} \;, \non
 && M_{\alpha i} \equiv 
   \sqrt{i+M-1 \over i+M} m_{i+M-1} 
 - \sqrt{\alpha-1\over \alpha} m_{\alpha - 1} 
 + \sum_{a = \alpha}^{i+M-1} {m_a \over \sqrt {a(a+1)}} \;,                
 \label{superpotGr}
\eeq
 where we have set $m_0 \equiv 0$. 
For the case of $M=1$ ($\alpha=1$), 
 this reduces to Eq.~(\ref{superpotCP}) for 
 $T^* {\bf C}P^{N-1}$. From
 the superpotential (\ref{superpotGr}), 
 its derivatives with respect to the fields are
\beq
 \del_{\ph_{i \alpha} } W 
   = - M_{\alpha i} \psi_{\alpha i} \;,\hs{5} 
 \del_{\psi_{\alpha i} } W 
   = - M_{\alpha i} \ph_{i \alpha} \; \hs{5} (\mbox{no sum}) \;.
\eeq
Therefore the origin of these coordinates, $\ph = \psi^T = 0$,
 is a vacuum, and this is the only one vacuum in 
 the finite region of these coordinates 
 where the metric is regular.
This model contains vacua as many as the coordinate patches, 
 like the $T^*{\bf C}P^{N-1}$ case.
In the first coordinate patch, 
 we have chosen the first $M$ row 
 vectors in $\Phi$ the unit matrix as in 
 Eqs.~(\ref{fixing1}) or (\ref{fixing2}). 
The other coordinate patches are given by 
 the other choices of gauge fixing conditions 
 making the other sets of $M$ row vectors 
 in $\Phi$ the unit matrix.
The number of such coordinate systems is 
 $_N C_M = N!/[M!(N-M)!]$. 
They are independent and 
 enough to cover the whole manifold, 
 so this model has $N!/[M! (N-M)!]$ vacua.
This number is invariant  
 under the duality between 
 $U(M)$ and $U(N-M)$ gauge groups.
It also reduces correctly to $N$ for $T^*{\bf C}P^{N-1}$ 
 when $M=1$ or $M=N-1$.

As in the $T^* {\bf C}P^{N-1}$ case, 
 we can understand the vacua of $T^* G_{N,M}$ 
 without local coordinates.
A point in the base $G_{N,M}$ corresponds to 
 an $M$-dimensional complex plane 
 through the origin in ${\bf C}^N$.
The vacua found above correspond to  
 mutually orthogonal $M$-planes spanned by
 arbitrary $M$ sets of axes chosen from the $N$ axes. 
Therefore the total number of vacua 
 is $_N C_M = N!/[M!(N-M)!]$. 
Since the $M$-planes of vacua are invariant under 
 $U(1)^{N-1}$ generated by $H_a$, 
 the vacua are fixed points.

Taking basis (\ref{basis}) in ${\bf C}^N$, 
 a point in $G_{N,M}$ 
 expressed by an $M$-plane in ${\bf C}^N$ can be 
 spanned by $M$ set of unit vectors 
\beq
  (e_i)' = U e_i \;, \hs{5} 
\eeq
 where $i = 1,\cdots,N-M$ and 
 $U$ is an unitary matrix, $U \in U(N)$. 
The vacua of mutually orthogonal $M$-planes are 
 spanned by arbitrary $M$ sets of 
 basis among orthogonal $N$-basis, 

The duality becomes manifest in this framework. 
We can represent a point in $G_{N,M}$ by 
 an $(N-M)$-plane complement to an $M$-plane.

\medskip
\underline{Example: the cotangent bundle over the Klein quadric}. 
An example is given for the Klein quadric $T^* G_{4,2} = T^* Q^4$ 
 ($N=4$ and $M=2$). 
There exist six coordinate systems $\ph^{(A)}_{i\alpha}$ 
 ($A=1,\cdots,6$) for the base manifold 
 corresponding to six choices of gauge fixing, 
 given by
\beq
 \Phi &=& 
 \left(
 \begin{array}{cc} 1 & 0 \cr 0 & 1 \cr 
         \ph^{(1)}_{11} & \ph^{(1)}_{12} \cr 
         \ph^{(1)}_{21} & \ph^{(1)}_{22}\end{array} 
 \right) \,, \hs{3}
 \left(
 \begin{array}{cc} 1 & 0 \cr \ph^{(2)}_{11} & \ph^{(2)}_{12} \cr 
           0 & 1 \cr \ph^{(2)}_{21} & \ph^{(2)}_{22}\end{array} 
 \right) \,, \hs{3}
 \left(
 \begin{array}{cc} 1 & 0 \cr \ph^{(3)}_{11} & \ph^{(3)}_{12} \cr 
           \ph^{(3)}_{21} & \ph^{(3)}_{22} \cr 0 & 1\end{array} \,,
 \right) \hs{3} \non
&& 
 \left(
 \begin{array}{cc} \ph^{(4)}_{11} & \ph^{(4)}_{12} \cr 1 & 0 \cr 
             0 & 1 \cr \ph^{(4)}_{21} & \ph^{(4)}_{22}\end{array} 
 \right)\,, \hs{3}
 \left(
 \begin{array}{cc} \ph^{(5)}_{11} & \ph^{(5)}_{12} \cr 1 & 0 \cr 
           \ph^{(5)}_{21} & \ph^{(5)}_{22} \cr 0 & 1\end{array}
 \right)  \,, \hs{3} 
 \left(
 \begin{array}{cc} \ph^{(6)}_{11} & \ph^{(6)}_{12} \cr 
           \ph^{(6)}_{21} & \ph^{(6)}_{22} \cr 1 & 0 \cr 0 & 1\end{array}
 \right) \, . 
 \label{Klein}
\eeq
Together with corresponding coordinates $\psi^{(A)}_{\alpha i}$ 
 for the cotangent space in Eq.~(\ref{fixing1}), 
 these six sets of coordinate systems are 
 enough to cover the whole manifold.
Therefore this model has the six vacua given by
\beq
 \left<\Phi\right> = 
 \left(
 \begin{array}{cc} 1 & 0 \cr 0 & 1 \cr 0 & 0 \cr 0 & 0 \end{array}
 \right) \,, \hs{3}
 \left(
 \begin{array}{cc} 1 & 0 \cr 0 & 0 \cr 0 & 1 \cr 0 & 0\end{array}
 \right)  \,, \hs{3}
 \left(
 \begin{array}{cc} 1 & 0 \cr 0 & 0 \cr 0 & 0 \cr 0 & 1\end{array}
 \right)  \,, \hs{3}
 \left(
 \begin{array}{cc} 0 & 0 \cr 1 & 0 \cr 0 & 1 \cr 0 & 0\end{array}
 \right)  \,, \hs{3}
 \left(
 \begin{array}{cc} 0 & 0 \cr 1 & 0 \cr 0 & 0 \cr 0 & 1\end{array}
 \right)  \,, \hs{3} 
 \left(
 \begin{array}{cc} 0 & 0 \cr 0 & 0 \cr 1 & 0 \cr 0 & 1\end{array}
 \right) \,, \label{vac-Klein}
\eeq
which are the origins of (\ref{Klein}) respectively, 
 with $\left<\Psi\right>=0$.
A set of two column vectors in each matrix in Eq.~(\ref{vac-Klein}) 
 is a set of orthogonal basis $e_i$ chosen from the four basis.

\medskip
In the case of $b \neq 0$, 
the superpotential (\ref{superpot2}) is 
\beq
 W &=& b \sum_{a=1}^{N-1} \sum_{n=0}^{\infty} (-1)^n  
    m_a \tr \left[ H_a \left(
     \begin{array}{cc} (\psi\ph)^n     & (\psi\ph)^n \psi \cr
               \ph (\psi\ph)^n & (\ph \psi)^{n+1} \end{array}
     \right)  \right] \non
 &=& b \sum_{a=1}^{N-1} \sum_{n=0}^{\infty} (-1)^n  
    m_a \tr \left[ H_a \left(
     \begin{array}{cc} (\psi\ph)^n & 0 \cr
                         0 & (\ph \psi)^{n+1} \end{array}
     \right)  \right] \;, \label{superpot3}
\eeq
where the last equality holds because $H_a$ are diagonal.
Similarly to the $T^*{\bf C}P^{N-1}$ case, 
 the origin $\ph = \psi^T =0$ of each patch is a vacuum 
 and we cannot have any other vacua.

\section{Massive HK quotient by $SU(M)$ gauge group}
In this section, we construct the massive HK NLSM 
with the $SU(M)$ gauge group. We eliminate the vector multiplets 
in the superfield formalism 
and find that this model does not have discrete vacua.

\subsection{Massive HK NLSM by $SU$ gauge group}
In this subsection, we consider ${\cal N}=2$ SUSY QCD with 
$N$-flavors and the $SU(M)$ gauge group.
We take the same matter field contents with $T^* G_{N,M}$ 
but gauge multiplets take values in the Lie algebra of $SU(M)$: 
$V= V^A T_A$ and $\Sigma = \Sigma^A T_A$ 
with $T_A$ generators of $SU(M)$.
Then the Lagrangian is given by 
\beq
&& {\cal L} = \int d^4 \theta
 \left[ \tr (\Phi\dagg\Phi e^V )  
 + \tr (\Psi\Psi\dagg e^{-V}) \right]  
 \non &&\hs{5} 
 + \left[ \int d^2\theta \,
       \left(\tr ( \Sigma \Psi \Phi ) 
      + \sum_{a=1}^{N-1} m_a \tr (\Psi H_a \Phi)\right) 
         + {\rm c.c.}\right] .
\label{linear-SU}
\eeq
We do not have any FI parameters 
because of the absence of 
any $U(1)$ gauge symmetry. 
The $SU(M)$ gauge transformation 
is given by the same way as in the $U(M)$ case 
and it is complexified to $SU(M)^{\bf C} = SL(M,{\bf C})$.
This model has an additional $U(1)_{\rm D}$ flavor symmetry, 
\beq
 \Phi \to \Phi' = e^{i \lam} \Phi \;, \hs{5}
 \Psi \to \Psi' = e^{- i \lam} \Psi , \label{U(1)D}
\eeq
which was gauged in $U(M)$ case.

We eliminate all auxiliary superfields 
in the superfield formalism.
Equations of motion for $V$, $\Sigma$ imply
\beq
 && \Phi\dagg\Phi e^V - e^{-V} \Psi\Psi\dagg = C {\bf 1}_M 
     \; \label{EOM-V2} , \\ 
 && \Psi \Phi = B {\bf 1}_M \;, \label{EOM-sigma2} 
\eeq
respectively, 
with $C(x,\theta,\thb)$ and $B(x,\theta,\thb)$ being 
vector and chiral superfields in 
the ${\cal N}=1$ superfields formalism.

The gauge field $V$ can be solved in 
terms of the dynamical fields from Eq.~(\ref{EOM-V2}) 
as
\beq
 e^V = {1 \over 2} (\Phi\dagg\Phi)^{-1} 
  \left(C {\bf 1}_M  \pm \sqrt {C^2 {\bf 1}_M 
     + 4 \Phi\dagg\Phi \Psi\Psi\dagg}\right) \,. 
  \label{sol-V2}
\eeq 
Since the equation $\det e^V =1$ holds, 
we get the equation 
\beq
  \det \left(C {\bf 1}_M \pm \sqrt{C^2 {\bf 1}_M  
              + 4 \Phi\dagg\Phi \Psi\Psi\dagg} \right) 
   =  2^M \det (\Phi\dagg \Phi) \;\label{c}
\eeq
which enables us to express $C$ in terms of dynamical fields 
implicitly: $C = C(\Phi,\Phi\dagg;\Psi,\Psi\dagg)$. 
On the other hand, Eq. (\ref{EOM-sigma2}) implies
\beq
 B = \1{M} \tr (\Phi \Psi) \;. \label{b}  
\eeq

Substituting the solution (\ref{sol-V2}) 
back into the Lagrangian
(\ref{linear-SU}), we obtain the K\"ahler potential 
\begin{equation}
 K = \pm \tr \sqrt{C^2 (\Phi,\Phi\dagg;\Psi,\Psi\dagg) {\bf 1}_M 
     + 4 \Phi\dagg\Phi \Psi\Psi\dagg} 
  \;, \label{kahler-SU}
\end{equation}
with $C$ satisfying the constraint (\ref{c}).
We should choose the plus sign
for the positivity of the metric.

Let us fix the complex gauge symmetry 
$SU(M)^{\bf C} = SL(M,{\bf C})$ 
to express the Lagrangian
in terms of independent superfields. 
We can take the similar gauge as the $b\neq 0$ case 
in $T^* G_{N,M}$:
\beq
 \Phi = \sig 
        \left(\begin{array}{c} {\bf 1}_M \cr \ph 
        \end{array}\right)P\;, \hs{5} 
 \Psi = P ({\bf 1}_M, \psi) \rho \;, \hs{5}
 P = ({\bf 1}_M + \psi\ph)^{-\1{2}} \;, \label{fixing-SU}
\eeq
with $\ph$ and $\psi$ being 
$[(N-M)\times M]$- and $[M\times (N-M)]$-matrix chiral 
superfields, respectively. 
Here, $\sig$ and $\rho$ are chiral superfields 
satisfying $\sig \rho = B$ from Eq.~(\ref{b}). 
We can consider $\sig$ and $\rho$ 
independent fields among these three fields. 

Substituting Eq.~(\ref{fixing-SU}) into 
the \kahler potential (\ref{kahler-SU}), 
we obtain the \kahler potential in terms of independent fields 
$\ph,\psi,\rho,\sigma$ and their conjugates. 
The superpotential also can be calculated as
\beq
 W = \sum_a m_a \sigma \rho \; 
   \tr \left[
    H_a \left(\begin{array}{c} {\bf 1}_M \cr \ph \end{array} 
    \right)
    ({\bf 1}_M + \psi\ph)^{-1}
    ({\bf 1}_M, \psi)  
   \right] \;. \label{superpot-SU}
\eeq

This target manifold has the isometry of 
 $U(N) =SU(N) \times U(1)_{\rm D}$, 
 in which the $SU(N)$ part is the same with $T^* G_{N,M}$. 
The \kahler potential does not receive 
 the \kahler transformation. 
As for the symmetry of the Lagrangian, 
 the superpotential is invariant under 
 the $U(1)$ fiber symmetry originated from (\ref{U(1)D})
\beq
 \sig \to \sig' = e^{i \lam} \sig \;, \hs{5} 
 \rho \to \rho' = e^{- i \lam} \rho \;,
\eeq
 besides the $U(1)^{N-1}$ symmetry of the massive 
 $T^* G_{N,M}$ model. 
Gauging this $U(1)_{\rm D}$ symmetry,
 we obtain the $T^*G_{N,M}$ model. 
Gauging $U(1)_D$ symmetry implies putting $B$ and $C$ in
 the constraints (\ref{EOM-V2}) and (\ref{EOM-sigma2}) 
 as constants and the constraints then become
 $T^*G_{N,M}$'s ones (\ref{EOM-V}) and (\ref{EOM-sig}), respectively.
This clarifies the bundle structure: 
 the set of $\sig$ and $\rho$ 
 is a fiber of quaternion with the total manifold being 
 the (quaternionic) line bundle 
 over $T^* G_{N,M}$.

\subsection{Vacua of $SU$ gauge theories}

We look for the vacua of the HK NLSM by the $SU$ gauge group.
The superpotential (\ref{superpot-SU}) of this model 
can be rewritten as
\beq
 W = \sigma \rho \sum_{a=1}^{N-1} \sum_{n=0}^{\infty} (-1)^n  
    m_a \tr \left[ H_a \left(
     \begin{array}{cc} (\psi\ph)^n & 0 \cr
                         0 & (\ph \psi)^{n+1} \end{array}
      \right)
      \right]
  \equiv \sigma \rho W_U \;,
\eeq
where $W_U$ (times $b$) denotes the superpotential 
(\ref{superpot2}) or (\ref{superpot3}) of 
the $U(M)$ gauge group with $b \neq 0$.
The derivatives of the superpotential with respect to fields 
are given by
$\del_{\psi} W = \sig \rho \del_{\psi} W_U$, 
$\del_{\ph} W = \sig \rho \del_{\ph} W_U$, 
$\del_{\rho} W = \sig W_U$ and $\del_{\sig} W = \rho W_U$. 
The vacuum condition is given by $\sig = \rho =0$, 
since $\del W_U = 0$ holds only at $\ph=\psi^T=0$ from 
the discussion in the last section, 
but $W_U \neq 0$ there.
Therefore this model has no discrete vacua, 
and so we cannot expect any wall solutions.

\section{BPS equation and its solution}
In this section, 
 we construct the BPS domain wall 
 in $N=2$ and $M=1$ case of $T^*G_{N,M}$
 i.e., $T^*{\bf C}P^1$.
In what follows, we consider $b\neq 0$ and $c=0$ case.
We assume that there exists domain wall solution 
 perpendicular to $y=x^2$ direction.
BPS domain wall solution is derived from vanishing of
 the SUSY transformation for fermions
\beq
0
 = i \sqrt 2 \sigma^\mu {\bar{\epsilon}} \partial_\mu \Phi^i 
 + \sqrt 2 \epsilon F^i \;
\eeq
with half SUSY condition
$
 e^{i\alpha} \sig^2 \bar\epsilon =i \epsilon 
$
where $e^{i\alpha}$ is a phase factor, $\Phi^i$ and $F^i$ are scalar
 and auxiliary fields, respectively.
In the case we consider now, the scalar field is given by
 $\Phi^i=\sqrt{\frac{b}{1+\varphi\psi}}
 \left(\begin{array}{c} 1 \\ \varphi \end{array}\right)$
 from Eq. (\ref{fixing2}).
Eliminating the auxiliary fields, the BPS equations are given by
\beq
 \del_2 \ph^i = - e^{i\alpha} 
g^{ij^*} \del_{j^*} W^* \;,  
\eeq
where $g^{ij^*}$ is inverse of the metric 
 $g_{ij^*}=\partial_i\partial{_{j^*}}K$ and $K$ is given
 by (\ref{kahler}) with (\ref{fixing2}).
Substituting the metric and the superpotential (\ref{superpot2}), 
 these BPS equations reduce to 
\beq
 && \del_2 \varphi = e^{i\alpha}  {m^* \over 4 b} K (1 + \varphi\psi)^2 
  \left[{|1+\varphi\psi|^2 + (1 +|\varphi|^2)(1+|\psi|^2) 
        \over |1+\varphi\psi|^2 (1+|\psi|^2)^2} \psi^*  
  + { (\varphi-\psi^*)^2 \varphi^* \over |1+\varphi\psi|^2 
  (1+|\varphi|^2)(1+|\psi|^2) } \right], \non
 && \del_2 \psi =  e^{i\alpha}  {m^* \over 4 b} K (1 + \varphi\psi)^2 
  \left[{|1+\varphi\psi|^2 + (1 +|\varphi|^2)(1+|\psi|^2) 
        \over |1+\varphi|^2 (1+|\varphi|^2)^2} \varphi^*  
      + {(\psi-\varphi^*)^2 \psi^* \over |1+\varphi\psi|^2 
      (1+|\varphi|^2)(1+|\psi|^2) } \right],
 \non
\label{eq:BPS_vw}
\eeq 
where $m$ is a mass parameter.
Now we must choose the phase $e^{i\alpha}$ to absorb 
 the phase of the parameter \footnote{
 For simplicity, we choose $m$ to be real positive 
 in the following.}
$m^*/b$ 
\begin{equation}
e^{i\alpha}{m^* \over b}=\left|{m \over b}\right|. 
\end{equation}
By subtracting the complex conjugate of the second equation from 
the first one in Eq.(\ref{eq:BPS_vw}), we obtain 
\begin{eqnarray}\label{eq:BPS_v-w*}
 &\!\!\! &\!\!\!
{\partial (\varphi-\psi^*) \over \partial y}=
 \left|{m \over b}\right|{K \over 4}
 \left[
 \left\{\left({1+\varphi\psi \over |1+\varphi\psi|}\right)^2 \varphi^*
-\left({1+\varphi^*\psi^* \over |1+\varphi\psi|}\right)^2\psi\right\}
 {(\varphi-\psi^*)^2 \over (1+|\varphi|^2)(1+|\psi|^2)}
\right.
 \\
 &\!\!\! 
+
 &\!\!\!
\left.
\left\{\left({1+\varphi\psi \over |1+\varphi\psi|}\right)^2{\psi^* \over 
 (1+|\psi|^2)^2}
-\left({1+\varphi^*\psi^* \over |1+\varphi\psi|}\right)^2{\varphi \over 
 (1+|\varphi|^2)^2}\right\}
\left\{|1+\varphi\psi|^2+(1+|\varphi|^2)(1+|\psi|^2)\right\}
\right], 
\nonumber
\end{eqnarray}
 whose right-hand side vanishes for $\varphi=\psi^*$. 
The BPS equation (\ref{eq:BPS_v-w*}) 
 dictates that $\varphi=\psi^*$ is valid for arbitrary $y$, 
 if an initial condition $\varphi=\psi^*$ is chosen at some $y$. 
Since we can choose the initial condition $\varphi=\psi^*$ at $y=-\infty$, 
 we find the BPS equations (\ref{eq:BPS_vw}) simply reduce to 
\beq
 \del_2 \varphi =   |m| \varphi \;,
\eeq
which is the BPS equation on the submanifold ${\bf C}P^1$
 defined by $\varphi = \psi^*$ \cite{ANNS}. 
Therefore we obtain a BPS wall configuration connecting two vacua 
 $\varphi=\psi^*=0$ at $y=-\infty$ to $\varphi=\psi^*=\infty$ at 
 $y=\infty$ along $\varphi=\psi^*$ with a constant phase $e^{i\phi_0}$ 
\beq
 \varphi = \psi^* = e^{|m| (y+y_0)}e^{i\phi_0}\;,    
 \label{solution} 
\eeq
where $y_0$ is also a constant representing the position of the wall. 
Thus we find two collective coordinates (zero modes) corresponding 
to the spontaneously broken translation ($y_0$) and $U(1)$ symmetry 
($\phi_0$). 

We can show that BPS solution (\ref{solution}) coincides with 
 that derived in component formalism \cite{GTT1}
 through the following field redefinition
 $\varphi\rightarrow X, \phi$
\begin{eqnarray}
 \varphi \equiv e^{u + i \phi}, 
\qquad 
  X=|b|\tanh u,
\end{eqnarray}
where $u,~\phi$ and $X$ are real scalar fields.
After the field redefinition, the theory of massive ${\bf C}P^1$
 model is described by $X$ and $\phi$, and 
 the wall solution (\ref{solution}) is mapped to
\begin{eqnarray}
 X=|b|\tanh |m|(y + y_0), \qquad 
\phi = \phi_0. \label{GPT-sol}
\end{eqnarray}
This solution coincides with that derived in 
 Ref.~\cite{GTT1}.

\section{Conclusion}
We have constructed massive NLSMs on cotangent bundle over
 Grassmann manifold 
 $T^*G_{N,M}$ and its generalization, the line bundle
 over $T^*G_{N,M}$ manifold in ${\cal N}=1$ superfield
 formalism with quotient method.
It was found that the former contains $N!/[M!(N-M)!]$ vacua while the
 latter has no discrete vacua.

The BPS wall solution was given in $N=2$ and $M=1$ case
 of $T^*G_{N,M}$ model, which corresponds
 to the Eguchi-Hanson manifold. 
More interesting case is $N=4$ and $M=2$ case since it is the simplest
 manifold other than $T^*{\bf C}P^{N-1}$.
The theory has six discrete vacua and it is expected that the theory
 has various interesting wall solutions, their junction and lump.

\vspace{1.0cm}

\noindent{\Large \bf Acknowledgements}\\

\noindent 
We thank Masashi Naganuma for a collaboration 
 in early stages of this work. 
M.~A. is grateful to the organizers in 
 SYM-PHYS10.
This work is supported in part by Grant-in-Aid 
 for Scientific Research from the Japan Ministry 
 of Education, Science and Culture  13640269 (NS). 
The work of M.~N. was supported by the U.~S. Department
 of Energy under grant DE-FG02-91ER40681 (Task B).

\end{document}